\documentclass[times,final]{elsarticle}

\usepackage{dib}
\usepackage{framed,multirow}

\usepackage{amssymb}
\usepackage{latexsym}
\usepackage[artemisia]{textgreek} 
\usepackage{listings} 

\usepackage{url}
\usepackage{xcolor}
\definecolor{newcolor}{rgb}{.8,.349,.1}

\usepackage[utf8]{inputenc} 
\DeclareTextSymbolDefault{\softL}{T1}
\DeclareTextSymbolDefault{\softl}{T1}
\DeclareTextCommand{\softL}{T1}{\v{L}}
\DeclareTextCommand{\softl}{T1}{\v{l}}
\DeclareUnicodeCharacter{013D}{\softL}
\DeclareUnicodeCharacter{013E}{\softl}    

\usepackage{longtable}
\usepackage[colorlinks]{hyperref}

\journal{Data in Brief}

\begin{document}

\lstset{columns=flexible, basicstyle=\ttfamily}
\verso{Ji\v{r}\'{i} Filipovi\v{c} \textit{et al}}

\begin{frontmatter}

\dochead{Data Article}
\title{Searching CUDA code autotuning spaces with hardware performance counters: data from benchmarks running on various GPU architectures}%

\author[ics]{Ji\v{r}\'{i} Filipovi\v{c}\corref{mycorrespondingauthor}}
\ead{fila@mail.muni.cz}
\cortext[mycorrespondingauthor]{Corresponding author}
\author[ics]{Jana Hozzov\'{a}}
\ead{hozzova@mail.muni.cz}
\author[ics]{Amin Nezarat}
\ead{aminnezarat@mail.muni.cz}
\author[ics]{Jaroslav O\softl{}ha}
\ead{348646@mail.muni.cz}
\author[ics]{Filip Petrovi\v{c}}
\ead{fillo@mail.muni.cz}
\address[ics]{Institute of Computer Science, Masaryk University, Botanick\'{a} 68a, 60200 Brno, Czech Republic}

\begin{abstract}

\noindent
We have developed several autotuning benchmarks in CUDA that take into account performance-relevant source-code parameters and reach near peak-performance on various GPU architectures. We have used them during the development and evaluation of a novel search method for tuning space proposed in \cite{Filipovic2021}. With our framework Kernel Tuning Toolkit, freely available at Github, we measured computation times and hardware performance counters on several GPUs for the complete tuning spaces of five benchmarks. These data, which we provide here, might benefit research of search algorithms for the tuning spaces of GPU codes or research of relation between applied code optimization, hardware performance counters, and GPU kernels' performance.

Moreover, we describe the scripts we used for robust evaluation of our searcher and comparison to others in detail. In particular, the script that simulates the tuning, i.e., replaces time-demanding compiling and executing the tuned kernels with a quick reading of the computation time from our measured data, makes it possible to inspect the convergence of tuning search over a large number of experiments. These scripts, freely available with our other codes, make it easier to experiment with search algorithms and compare them in a robust way.

During our research, we generated models for predicting values of performance counters from values of tuning parameters of our benchmarks. Here, we provide the models themselves and describe the scripts we implemented for their training. These data might benefit researchers who want to reproduce or build on our research.


\end{abstract}


\end{frontmatter}


\section*{Data Article template}

{\fontsize{7.5pt}{9pt}\selectfont
\noindent\textbf{Specifications Table} 

\begin{longtable}{|p{33mm}|p{94mm}|}
\hline
\endhead
\hline
\endfoot
Subject                & Computer Science \\
\hline                         
Specific subject area  & Auto-tuning GPU kernels using hardware performance counters\\
\hline
Type of data     & Tables\newline
                         Python and R scripts
                        \\       
\hline
How data were acquired &

\textbf{Raw autotuning data:} Using our autotuning framework KTT, we measured computation time and collected hardware performance counters for whole tuning spaces of five benchmark CUDA codes on four GPUs. Kernel Tuning Toolkit (KTT) is freely available in Github repository \url{https://github.com/HiPerCoRe/KTT}, five benchmarks are also there in 'examples' folder. These benchmarks cover a wide range of computational problems: computing convolution, Coulomb summation in three dimensions, matrix multiplication, matrix transposition and N-body problem. They also differ in sizes of their tuning spaces. \newline
\textbf{Prediction models:} Using our scripts, also available in KTT repository at Github, we trained models with the raw tuning data.
\\
\hline                         
Data format            &
                         Raw\newline
                         Analyzed\newline
                         Scripts
                         \\
\hline                         
Parameters for         
data\newline 
collection             &
\textbf{Raw autotuning data:} Computation time and performance counters were measured for five benchmarks (GEMM, Convolution, Matrix transposition, 3D Coulomb summation and n-body) bundled with KTT, git tag \texttt{v1.3-profile-searcher}. We ran them on four GPUs: GeForce GTX 680, GeForce GTX 750, GeForce GTX 1070 and GeForce RTX 2080. KTT was configured to perform exhaustive exploration of tuning spaces on each GPU under our test with profiling switched on. Size of input for each benchmark was chosen so that the kernel execution took 1-10 milliseconds. For the GEMM benchmark, data for several input sizes were collected on GTX GeForce 1070.  
\newline
\textbf{Prediction models:} Prediction models were trained for all hardware performance counters, local and global size. 
\\
\hline
Description of          
data\newline 
collection             &
\textbf{Raw autotuning data:} KTT performed exhaustive exploration of complete tuning spaces (sets of all executable tuning configurations) of tested benchmarks for each GPU. Each tuning configuration contains information about tuning parameters (affecting how GPU kernel is created and executed), the runtime of the kernel and hardware performance counters provided by NVIDIA CUPTI library. Tuning configurations which cannot be executed on a particular GPU are not stored.
\newline
\textbf{Prediction models:} Models were created from the raw autotuning data with scripts \texttt{create\_nonlinear\_models.R} and \texttt{generate-knowledge-base.py} available with profile-based searcher in KTT.
\\
\hline                         
Data source location   & 
                         Institute of Computer Science, Masaryk University\newline
                         Brno\newline
                         Czech Republic\newline
                         49.211N, 16.598E
\\
\hline                         
\hypertarget{target1}
{Data accessibility}   & 
                         Repository name: Mendeley Data\newline
                         Data identification number: doi:10.17632/nn53dskr7z.1\newline
                         Direct URL to data: http://dx.doi.org/10.17632/nn53dskr7z.1\newline
\\                         
\hline                         
Related                 
research\newline
article                & 
                         Filipovi\v{c}, J., Hozzov\'{a}, J., Nezarat, A., O\softl{}ha, J., Petrovi\v{c}, F., Using hardware performance counters to speed up autotuning convergence on GPUs, Future Generation Computer Systems. In Press.
\end{longtable}
}

\section*{Value of the Data}


\begin{itemize}
\itemsep=0pt
\parsep=0pt
\item 
Raw autotuning data contain, to the best of our knowledge, the first freely available complete tuning spaces of several CUDA kernels prepared for autotuning, alongside their computation times and hardware performance counters measurements on several GPUs. Scripts make it easier to experiment with searching tuning space in a controlled environment, so the results of searchers are comparable. 
\item 
These data will help those researching how to search the tuning spaces of GPU codes or those interested in mining the data related to hardware performance counters and GPU kernels' performance.
\item 
With raw autotuning data, new search algorithms for navigating the tuning spaces can be easily evaluated for multiple GPUs (even those unavailable to the researchers), skipping high time demands of actually compiling, running and measuring. Moreover, the global optimum of the tuning space is known from data.  
\item With scripts for simulated and real-time tuning, the results of others (with a new search method or a new prediction model for performance counters) can be consistently compared to the results of our searcher.
\item Availability of KTT autotuner and scripts for model preparation allows users to expand our dataset by measurement on their own GPUs, or their own benchmarks.
\end{itemize}
\section*{Data Description}



\subsection*{Raw Autotuning Data}
Raw autotuning data were produced by Kernel Tuning Toolkit 1.3\footnote{\url{https://github.com/HiPerCoRe/KTT/releases/tag/v1.3-profile-searcher}} running on GPUs listed in Table~\ref{tab:gpus}. For each benchmark listed in Table~\ref{tab:benchmarks} (available in KTT repository in folder \texttt{examples} as cltune-conv, coulomb\_sum\_3d, cltune-gemm, mtran and nbody), the exhaustive search of the whole tuning space was executed, measuring computation time and hardware performance counters. For details on benchmarks and their tuning spaces, see~\cite{petrovic2020benchmark}. 

\begin{table}
\caption{GPU devices used to obtain our data.}
\label{tab:gpus}
\centering
\begin{tabular}{|l|l|l|l|}
\hline
Device                  & Architecture  & Released & Abbreviation\\ \hline
GeForce GTX 680  	& Kepler        & 2012     & 680 \\
GeForce GTX 750  	& Maxwell       & 2014     & 750 \\
GeForce GTX 1070 	& Pascal        & 2016     & 1070 \\
GeForce RTX 2080 	& Turing      	& 2018     & 2080 \\
\hline
\end{tabular}
\end{table}

\begin{table}
\caption{A list of the benchmarks used to obtain our data.}
\label{tab:benchmarks}
\centering
\begin{tabular}{|l|l|l|}
    \hline
    Benchmark 	& Description &  Abbreviation\\
    \hline
    Convolution	& 2D convolution kernel using $7 \times 7$ filter adopted from~\cite{nugteren2015cltune}. & conv  \\
    Coulomb 3D  & Direct coulomb summation on 3D lattice, introduced in~\cite{filipovic2017autotuning}. & coulomb\\
    GEMM	& Matrix-matrix multipication adopted from~\cite{nugteren2015cltune}, tuning space & gemm-reduced\\
		& reduced as in~\cite{nugteren2018clblast}.  & \\
    Transpose   & Out-of-place matrix transposition, adopted from NVIDIA & mtran\\
		& CUDA SDK 10.0. & \\
    N-body	& N-body kernel, adopted from NVIDIA CUDA SDK 10.0. & nbody \\
    \hline
\end{tabular}
\end{table}

The raw data are available at \url{http://dx.doi.org/10.17632/nn53dskr7z.1} in directory 'raw-autotuning-data'. They are stored as CSV files with naming convention containing the abbreviation of GPU, the abbreviation of benchmark, and suffix \texttt{\_output.csv}. For example, data obtained on GeForce GTX 1070 and N-body benchmark are stored in \\\texttt{1070-nbody\_output.csv}. There are special cases for GEMM benchmark, where we obtained data on small and highly-rectangular matrices. Those benchmarks are abbreviated as \texttt{1070-gemm-128-128-128} (multiplication of $128 \times 128$ matrices), \texttt{1070-gemm-16-4096-4096} (multiplication of matrix $16 \times 4096$ with matrix $4096 \times 16$), \texttt{1070-gemm-4096-16-4096} (multiplication of matrix $4096 \times 16$ with matrix $4096 \times 4096$) and \texttt{1070-gemm-4096-4096-16} (multiplication of matrix $4096 \times 4096$ with matrix $4096 \times 16$). However, those benchmarks are measured for GeForce GTX 1070 only.

The CSV files produced by Kernel Tuning Toolkit are formatted as follows:
\begin{itemize}
  \item the first line is the header containing names of columns;
  \item each other line contains the profile of one tuning configuration (a combination of tuning parameters, which produces unique CUDA kernel source code and execution setting);
  \item if some configuration cannot be executed on a given GPU (e.g., because of insufficient hardware resources), it is not included in CSV (therefore, the same benchmarks can produce CSV files with a different number of lines when executed on different GPUs).
\end{itemize}

Each line of the CSV file contains the following types of columns:
\begin{itemize}
  \item \textit{Kernel name}: the name of the benchmarked kernel (the same for one type of benchmark);
  \item \textit{Computation duration (\textmu s)}: the duration of the benchmarked kernel and unit the time is measure in;
  \item \textit{Global size} and \textit{Local size}: The global and local size of the executed kernel (number of threads and block size in CUDA terminology). The size is counted as a scalar number; it reflects an overall number of threads with no respect to the grid or block dimensionality;
  \item Tuning parameters: the benchmark-specific tuning parameters, named in capitals by our convention (e.g., \texttt{VECTOR\_TYPE} or \texttt{CR});
  \item Hardware performance counters: performance counters measured on particular GPU (e.g., \texttt{dram\_utilization} or \texttt{inst\_fp\_32}).
\end{itemize}
Please note that not all available hardware performance counters were measured due to time demand to measure the complete tuning space. The performance counters set differs from GPU to GPU because different architectures implement different performance counters. The biggest change is with GeForce RTX 2080, where the performance counters are completely re-designed and re-named. 

\begin{table}
\caption{Input sizes used for gathering raw tuning data. The shown number determines size of input matrix/matrices in both dimensions with conv, gemm-reduced and mtran benchmarks. With nbody benchmark, the shown number determines number of simulated bodies. Finally, with coulomb benchmark, the first number is size of a 3D grid (the same in all dimensions), whereas the second number determines the number of atoms.}
\label{tab:input-sizes}
\centering
\begin{tabular}{|l|rrrr|}
	\hline
           & 680 & 750 & 1070 & 2080\\
        \hline
  conv  	& 2048 & 4096 & 4096 & 4096 \\
  coulomb 	& 256, 64 & 256, 64 & 256, 64 & 256, 64 \\
  gemm-reduced 	& 1024 & 1024 & 1024 & 2048 \\
  mtran 	& 8192 & 8192 & 8192 & 8192 \\
  nbody 	& 16384 & 16384 & 16384 & 16384\\
  \hline
\end{tabular} 
\end{table}

Input size for kernels was selected so that the kernel execution took approximately 1 - 10 milliseconds. These sizes obviously differ for each benchmark and GPU, Table~\ref{tab:input-sizes} summarizes them.

\subsection*{Prediction Models for Performance Counters}
We provide pre-computed prediction models for performance counters. All predict the global size, the local size, and performance counters relevant for the GPU of the training raw autotuning data. For a detailed list of performance counters and implemented models, see \cite{Filipovic2021}.

\subsubsection*{Least-squares Nonlinear Models}
Nonlinear prediction models were produced with the script \texttt{create\_least\_squares\_models.R} bundled with KTT in 'profile-searcher/scripts-prep/'. For each raw tuning data file (i.e. for each benchmark and each GPU, and for all input sizes of gemm benchmark on GeForce GTX 1070), we ran the script, producing multiple models for each performance counter, each for a different combination of values of binary tuning parameters. Please see the section Experimental Design, Materials and Methods for details on how the models are generated. It might make it easier to understand the format of the model files. The least-squares nonlinear models are stored as CSV file, following a similar naming convention as raw autotuning data: the abbreviation of GPU, the abbreviation of benchmark and suffix \texttt{-model\_[number].csv}. The special cases for GEMM benchmark are named analogously as their raw tuning data files. 

The CSV files produced by the script contain three sections. The first section includes a line for each tuning parameter describing an expression for coding this parameter, as coded values of tuning parameters are used to predict the values of performance counters). The second section includes one line called \texttt{Condition} describing a logical condition of values of binary parameters this model was trained for. Furthermore, the third section includes a line for each performance counter, describing an expression for predicting the given performance counter's value with the coded values of tuning parameters. 

\subsubsection*{Decision Tree}
Decision-tree prediction models were produced with the script \texttt{generate\_decision\_tree\_model.py} bundled with KTT in 'profile-searcher/scripts-prep/'. The script takes raw tuning data as an input and creates a predictive model of performance counters. Please see the section Experimental Design, Materials and Methods for details on how the models are generated. The resulting decision tree is stored as a pickle file together with a CSV file containing a list of all performance counters predicted by the model.

\section*{Experimental Design, Materials and Methods}



\subsection*{Obtaining Raw Tuning Data via Kernel Tuning Toolkit}
The raw tuning data are obtained during an autotuning process performed by Kernel Tuning Toolkit with GPU of our interest. Note that we recommend using version tagged v1.3-profile-searcher, which couples KTT 1.3 with profile-based searcher and benchmarks listed in Table~\ref{tab:benchmarks} prepared for collecting raw data or executing tuning with the profile-based searcher. We can explore the tuning space of any benchmark bundled with KTT (in 'examples' folder) or user-provided code. To obtain tuning data with hardware performance counters, KTT has to be built with enabled profiling (e.g., by calling \texttt{premake5 --profiling=cupti gmake}, see KTT documentation for further details). Moreover, the profiling has to be switched on in the autotuned application by calling \texttt{ktt::Tuner::set\-Kernel\-Profiling()} method (the profiling in benchmarks bundled with KTT can be switched on by setting \texttt{USE\_PROFILING} macro to 1). 

KTT can explore either the entire tuning space (default behaviour) or only its subset. In the latter case, it is recommended to use random searcher to randomize the observed subset (using method \texttt{ktt::Tuner::setSearcher()}). After the search of the space is complete, the tuning data are stored in CSV files by method \texttt{ktt::Tuner::Print\-Result()}). All benchmarks bundled with KTT stores resulting CSV and can execute exhaustive exploration of the tuning space by setting preprocessor macro \texttt{EXHAUSTIVE\_SEARCH} to 1. For more information about KTT methods and implementation of new autotuned codes, we refer to its documentation~\footnote{\url{https://github.com/HiPerCoRe/KTT}}.

\subsection*{Generating Prediction Models from Raw Tuning Data}
We provide scripts to generate prediction models from raw tuning data. These scripts take tuning space of the problem with collected performance counters and train a model that predicts performance counters' values when given a tuning configuration. 

\subsubsection*{Generating Least-square Regression Non-linear Models}
We provide two scripts in 'profile-searcher/scripts-prep' folder in the Github KTT repository.

The main script \texttt{create\_least\_squares\_models.R} trains nonlinear models: it takes the tuning data, and for each performance counter, it generates a model that predicts its value based on values of tuning parameters. To increase the accuracy of such prediction, we divide the tuning space into subspaces based on values of binary tuning parameters, as we suspect these have a profound influence on the performance counters. Thus, we generate several models for each performance counter, each model applicable only for a given combination of values of binary tuning parameters. An example of its usage is shown in Listing~\ref{lst:create-nm}. We recommend R v3.4.4, no special Rcran libraries are necessary.

\begin{lstlisting}[label=lst:create-nm, caption=Generating nonlinear models for GEMM benchmark on GTX GeForce 1070]
Rscript ./create_least_squares_models.R 1070-gemm-reduced_output.csv \
        1070-gemm-reduced 4:19 2,3,19:62
\end{lstlisting}

It takes four arguments:
\begin{itemize}
 \item \texttt{[input file name]} e.g. 1070-gemm-reduced\_output.csv, must follow the formatting of raw tuning data, as described above in section Data Description;
 \item \texttt{[prefix for output files names]} e.g. 1070-gemm-reduced, this will be used to name output files with models, \texttt{-model\_[number].csv} will be added;
 \item \texttt{[numbers of columns with tuning parameters in input file]} in format allowing to set individual columns and interval of columns (in format 'from:to') e.g. 2,5:12 meaning columns 2 and 5 through 11, counting starts at 0;
 \item \texttt{[numbers of columns with performance counters in input file]} in the same format.
\end{itemize}

After parsing the script arguments and reading the input file, we code the tuning parameters' values, i.e., scale them to the range of $<-1,1>$. This step is recommended in any regression model design, as models generally do not work well with the absolute values of the factors. Next, we select the values of tuning parameters that will determine training data. In other words, we do not choose data points (rows from the input file) for training randomly. We select a few values of non-binary tuning parameters and then include all available combinations in the training dataset. We need to moderate the number of values to prevent an exponential increase in training data size or a poor sampling of some part of the tuning space due to constraints. 

The function takes two arguments: the formula and the training data. The formula includes the factors, i.e. coded tuning parameters, and arithmetic operations with them. To make the models nonlinear, we include multiplications of factors (to capture their interactions) and quadratic terms. The training data include rows from the input file with selected values of tuning parameters and corresponding values of the given profiling counter. 

The output of the script are multiple files named \texttt{[output\_name]-model\_[number].csv}. The number of models corresponds to the number of combinations of values of binary parameters. If a model cannot be created for a specific combination of binary parameters (e.g. there are no data due to constraints), the closest model (i.e., the minimal number of values of binary tuning parameters differs) fills in and is printed in the output file. The format and contents of model files are described in the above section Data Description.
 
The script \texttt{generate\_least\_squares\_models.py} makes it quick and easy to generate models for all our raw tuning data. It requires python3, with docopt library installed. It takes one argument, \texttt{--benchmark}. The option \texttt{--benchmark GPU} generates models for benchmarks conv, coulomb, gemm-reduced, mtran and nbody for all GPUs. Option \texttt{--benchmark GEMM} generates models for different input sizes of benchmark gemm-reduced on GeForce GTX 1070. Users may need to modify the script to accommodate the names of folders with data.

\subsubsection*{Generating Decision Trees}

The script \texttt{generate\_decision\_tree\_model.py} for decision-tree model preparation is stored in 'profile-searcher/scripts-prep' folder in a KTT distribution. While generating the model, users have to supply the CSV file containing explored tuning space, columns containing tuning parameters and profiling counters with parallelism configuration ("Global size" and "Local size" columns) in the same format as with least-square regression nonlinear models, see Listing~\ref{lst:create-kb}.

\begin{lstlisting}[label=lst:create-kb, caption=How to call script for generating nonlinear model]
python3 generate_decision_tree_model.py \
   -i 1070-gemm-reduced_output.csv -t 4:19 -c 2,3,19:62
\end{lstlisting}

The script builds models predicting performance counters using optimized decision trees. The performance counters prediction with decision trees is computationally more efficient than with least-squares model; therefore, we recommend them as a default choice. The decision trees have high precision in densely sampled tuning spaces, but they are poor in extrapolation. Therefore, if only a smaller part of the tuning space is sampled, we recommend testing whether the least-squares method would bring better precision and faster tuning convergence. 

The resulting decision tree is stored in files with suffix \texttt{DT.sav} and a list of hardware performance counters predicted by the script has suffix \texttt{DT.sav.pc}. For example, model for the file \texttt{1070-gemm-reduced\_output.csv} is stored in \texttt{1070-gemm-reduced\_output\_DT.sav}.

\subsection*{Execution of Simulated Tuning}
Simulated tuning scripts \texttt{simulated-profiling-searcher.py} perform a search of the autotuning space on a pre-computed tuning space. It requires auxiliary files \texttt{base.py} and \texttt{mlKTTPre\-dictor.py} distributed with KTT. It also requires python3, with installed libraries docopt, numpy, pandas, pickle and sklearn. 

Instead of real execution and profiling of autotuned kernels (obtaining their runtime and hardware performance counters), it reads stored raw autotuning data (i.e. just simulates their execution and profiling) and performs a search on them. The advantage of this approach is that it is performed much faster than real tuning (as no compilation, execution and profiling are performed); therefore, the simulated tuning experiments can be performed many times to get statistically relevant data. The simulated run also does not require installation of KTT and GPU we are simulating autotuning for. The convergence of search method is measured and can be compared in means of the number of search steps (equal to the number of kernel executions performed by KTT in a real environment).

Listing \ref{lst:simulated-tuning} shows three examples of running the script. We want to tune gemm-reduced benchmark. The models for predicting performance counters were trained on GeForce GTX 750, and we want to use them to guide the search on GeForce GTX 1070. These three commands differ in the model used. The first one does not predict anything, only reads the performance counters' values from the provided raw tuning data file. The second one uses a decision tree, and the third one uses least-squares nonlinear models.

\begin{lstlisting}[label=lst:simulated-tuning, caption=Example of simulated-profiling-searcher.py.]
python3  -W ignore ./simulated-profiling-searcher.py \
    -o 1070-gemm-reduced_output.csv --oc 6.1 --mp 15 --co 1920 \
    --cm 750-gemm-reduced_output.csv --ic 5.0 -p 1 -t 4:19 -c 2,3,19:62 \
    --compute_bound -e 1000 -i 1000
python3  -W ignore ./simulated-profiling-searcher.py \
    -o 1070-gemm-reduced_output.csv --oc 6.1 --mp 15 --co 1920 \
    --dt 750-gemm-reduced_output_DT.sav --ic 5.0 -p 1 \
    -t 4:19 -c 2,3,19:62 --compute_bound -e 1000 -i 1000
python3  -W ignore ./simulated-profiling-searcher.py \
    -o 1070-gemm-reduced_output.csv --oc 6.1 --mp 15 --co 1920 \
    --ls 750-gemm-reduced --ic 5.0 -p 1 -t 4:19 -c 2,3,19:62 \
    --compute_bound -e 1000 -i 1000

\end{lstlisting}

The script takes multiple arguments:
\begin{itemize}
 \item \texttt{-o [raw tuning data file]} the raw tuning data file following the format described in the section Data Description
 \item \texttt{--oc [compute capability of GPU used to produce raw tuning data]} e.g. 6.1, if raw tuning data came from GeForce GTX 1070 \footnote{Note that values of several arguments, such as compute capability of GPUs, number of multiprocessor or CUDA cores, and column indexes for computation time, tuning parameters and profiling counters are available in script \texttt{autobench.py} for our raw tuning data. This script is described later in this section.}
 \item \texttt{--mp [number of multiprocessors on that GPU]} e.g. 15, if raw tuning data came from GeForce GTX 1070
 \item \texttt{--co [number of CUDA cores on that GPU]} e.g. 1920, if raw tuning data came from GeForce GTX 1070
 \item one of the following 
 \begin{itemize}
   \item \texttt{--cm [raw tuning data file]} with this option, no prediction of values of performance counters is computed, their actual values are read from the file with given raw tuning data
   \item \texttt{--dt [decision tree model file]} with this option, decision tree model is employed to predict values of performance counters
   \item \texttt{--ls [prefix for least squares model files]} with this option, least squares nonlinear models are employed to predict values of performance counters
 \end{itemize}
 \item \texttt{--ic [compute capability of the GPU of training data for model]} e.g. 5.0, if model was trained with data from GeForce GTX 750
 \item \texttt{-p [column with computation time]} always 1 in provided raw tuning data
 \item  \texttt{-t [columns with TP]} 4:19 for gemm-reduced bechmark
 \item  \texttt{-c [columns with PC]} 2,3,19:62 for gemm-reduced benchmark on GeForce GTX 1070
 \item  \texttt{-e [number of experiments]} sets how many times the experiment is repeated to get more stable results in case of randomized searchers
 \item  \texttt{-i [number of iterations]} sets how many tuning iterations (i.e., search steps) are performed per experiment
 \item \texttt{--compute-bound} or \texttt{--memory-bound} e.g., \texttt{--compute-bound} as that is the character of gemm-reduced problem
\end{itemize}

For details on the algorithm of profile-based search, please see \cite{Filipovic2021}.

The results of the analysis are stored as CSV files of the following format. The first line contains a header with names of the columns. Each column presents an iteration of the searcher (i.e., the exploration of the next tuning configuration, requiring its profiling). The first column contains the iteration number. The second column contains an average runtime with the standard deviation of the best kernel known in this iteration when the random searcher is utilized. The third column contains an average runtime with the standard deviation of the best kernel known in this iteration when the profile-based searcher is utilized. 

The script \texttt{autobench.py} makes it easy to run simulated tuning for all our raw tuning data. It takes two arguments, \texttt{--benchmark} and \texttt{--method}. The option \texttt{--benchmark GPU} runs simulated tuning for benchmarks conv, coulomb, gemm-reduced, mtran and nbody for all GPUs. Option \texttt{--benchmark GEMM} runs simulated tuning for different input sizes of benchmark gemm-reduced on GeForce GTX 1070. The option \texttt{--method} has three possible arguments \texttt{Exact}, \texttt{DecisionTree} or \texttt{LeastSquares} to denote the used model for predicting the values of performance counters. Users may need to modify the script to accommodate the names of folders with data. Moreover, the script can be used as a source of information on possible values of several command-line arguments of \texttt{simulated-profiling-searcher.py}, such as compute capabilities of different GPUs, number of their multiprocessors and CUDA cores, and indexes of columns for computation time, tuning parameters and performance counters in raw tuning data we provide.

We used the simulated tuning to analyze the convergence speed of the profile-based searcher proposed in~\cite{Filipovic2021} and of the random search. During the analysis, the autotuning is performed in the defined number of iterations. In each iteration of the searching process, the runtime of the best kernel found is logged. The autotuning is performed multiple times, so we obtain an average speed of the best kernel for each iteration over multiple autotuning executions. 

Other search methods or modifications based on our profile-based searcher might be easily added to scripts and compared consistently.

\subsection*{Execution Real-time Tuning}
Real-time tuning performs a search of the autotuning space without a pre-computed tuning space. The autotuned kernels are actually compiled, executed and profiled during the search. This is far more demanding than the simulated tuning described above, but it makes it possible to measure the actual time per tuning search iteration. The convergence of search method is measured and can be compared in means of tuning time.

Real-time tuning can be executed by running a compiled benchmark. For benchmarks bundled with KTT, the preparation includes the following steps:
\begin{itemize}
 \item KTT needs to be compiled with profiling. i.e. \texttt{premake5 --profiling=cupti gmake} needs to be run before building it. In case of older GPU architectures, \\use \texttt{--profiling=cupti-legacy} instead.
 \item In the code of the benchmark (in \texttt{cpp} file), a preprocessor macro \texttt{EXHAUSTIVE\_SEARCH} has to be set to 0.
 \item The random search is used by default. To test profile-based searcher~\cite{Filipovic2021}, the macro \texttt{USE\_PROFILE\_SEARCHER} has to be set to 1 in the code of the benchmark (in \texttt{cpp} file).
 \item The time for autotuning is restricted to a certain value set by macro \texttt{TUNE\_SEC}. This time can be altered, for example, with \texttt{TUNE\_SEC 60} performs autotuning for 60 seconds.
 \item The prediction models needed for profile-based searcher have to be in 'KTT/profile-searcher\-/models' folder.
\end{itemize}

Listing \ref{lst:real-time-tuning} shows examples of running a single real-time tuning for each benchmark and saving the log file.

\begin{lstlisting}[label=lst:real-time-tuning, caption=Running a single execution of real-time tuning on every benchmark]
cd KTT/build/x86_64_Release/
./conv_cuda > conv_experiment_1.log
./coulomb_sum_3d_cuda > coulomb_experiment_1.log
./gemm_cuda > gemm_experiment_1.log
./mtran_cuda > mtran_experiment_1.log
./nbody_cuda > nbody_experiment_1.log
\end{lstlisting}

The benchmark executable has two arguments; however, both of them have default values:
\begin{itemize}
 \item \texttt{[platform index]} default value 0, cannot be changed when using CUDA
 \item \texttt{[device index]} default value 0, users may need to change this if multiple GPUs are available
\end{itemize}

The input size can be modified in the source code of the given benchmark, in the \texttt{main} function. Again, reasonable default values are set. In our evaluation in \cite{Filipovic2021}, we set the input size of benchmarks according to Table~\ref{tab:input-sizes}. 

Proper evaluation of searcher's convergence in means of tuning time requires multiple runs of real-time tuning. This can be easily done by executing the benchmark multiple times and generating a log file for each run. For easier processing of the results in multiple log files, we provide script \texttt{histogram.py}. See an example of its usage in Listings~\ref{lst:histogram}.

\begin{lstlisting}[label=lst:histogram, caption=Processing multiple runs of real-time tuning]
python3 ./histogram.py -s gemm-experiments -t 300 
\end{lstlisting}

It takes two arguments:
\begin{itemize}
 \item \texttt{-s [folder with log files from real-time tuning runs]};
 \item \texttt{-t [time]} in seconds that denotes the maximum running time that is analyzed.
\end{itemize}

Results of \texttt{histogram.py} script are stored as CSV files of the following format. The first line is a header containing names of the columns. Each row contains measurement in each second of the autotuning process. The missing row indicates that no new data was available in particular second of the tuning process (this may happen in the initial part of the tuning if the first profiled kernel runs a long time). The rows contain the following columns:
\begin{itemize}
 \item the time (in seconds) from the beginning of the autotuning;
 \item average runtime of the best kernel known in the corresponding time;
 \item standard deviation of the runtime;
 \item minimal runtime;
 \item maximal runtime.
\end{itemize}
The data are collected from all available executions of the autotuning (all log files in folder passed by argument \texttt{-s}).

\section*{Acknowledgments}
The work was supported from European Regional Development Fund-Project "CERIT Scientific Cloud" (No. CZ.02.1.01\-/0.0/0.0/16\_013/0001802). Computational resources were supplied by the project "e-Infrastruktura CZ" (e-INFRA LM2018140) provided within the program Projects of Large Research, Development and Innovations Infrastructures.

\section*{Declaration of Competing Interest}
The authors declare that they have no known competing
financial interests or personal relationships which have, or could be
perceived to have, influenced the work reported in this article. 

\bibliographystyle{model1-num-names}
\bibliography{refs}

\begin{thebibliography}{5}
\expandafter\ifx\csname natexlab\endcsname\relax\def\natexlab#1{#1}\fi
\providecommand{\url}[1]{\texttt{#1}}
\providecommand{\href}[2]{#2}
\providecommand{\path}[1]{#1}
\providecommand{\DOIprefix}{doi:}
\providecommand{\ArXivprefix}{arXiv:}
\providecommand{\URLprefix}{URL: }
\providecommand{\Pubmedprefix}{pmid:}
\providecommand{\doi}[1]{\href{http://dx.doi.org/#1}{\path{#1}}}
\providecommand{\Pubmed}[1]{\href{pmid:#1}{\path{#1}}}
\providecommand{\bibinfo}[2]{#2}
\ifx\xfnm\relax \def\xfnm[#1]{\unskip,\space#1}\fi
\bibitem[{Filipovi\v{c} et~al.(2021)Filipovi\v{c}, Hozzov\'{a}, Nezarat,
  O\softl{}ha, and Petrovi\v{c}}]{Filipovic2021}
\bibinfo{author}{J.~Filipovi\v{c}}, \bibinfo{author}{J.~Hozzov\'{a}},
  \bibinfo{author}{A.~Nezarat}, \bibinfo{author}{J.~O\softl{}ha},
  \bibinfo{author}{F.~Petrovi\v{c}},
\newblock \bibinfo{title}{Using hardware performance counters to speed up
  autotuning convergence on {GPU}s},
\newblock \bibinfo{journal}{Future Generation Computer Systems}
  \bibinfo{volume}{In Press} (\bibinfo{year}{2021}).
\bibitem[{Petrovi\v{c} et~al.(2020)Petrovi\v{c}, St\v{r}el\'{a}k, Hozzov\'{a},
  O\softl{}ha, Trembeck\'{y}, Benkner, and
  Filipovi\v{c}}]{petrovic2020benchmark}
\bibinfo{author}{F.~Petrovi\v{c}}, \bibinfo{author}{D.~St\v{r}el\'{a}k},
  \bibinfo{author}{J.~Hozzov\'{a}}, \bibinfo{author}{J.~O\softl{}ha},
  \bibinfo{author}{R.~Trembeck\'{y}}, \bibinfo{author}{S.~Benkner},
  \bibinfo{author}{J.~Filipovi\v{c}},
\newblock \bibinfo{title}{A benchmark set of highly-efficient {CUDA} and
  {OpenCL} kernels and its dynamic autotuning with kernel tuning toolkit},
\newblock \bibinfo{journal}{Future Generation Computer Systems}
  \bibinfo{volume}{108} (\bibinfo{year}{2020}) \bibinfo{pages}{161--177}.
\bibitem[{Nugteren and Codreanu(2015)}]{nugteren2015cltune}
\bibinfo{author}{C.~Nugteren}, \bibinfo{author}{V.~Codreanu},
\newblock \bibinfo{title}{{CLTune}: A generic auto-tuner for {OpenCL} kernels},
\newblock in: \bibinfo{booktitle}{Proceedings of the IEEE 9th International
  Symposium on Embedded Multicore/Many-core Systems-on-Chip ({MCSoC})},
  \bibinfo{year}{2015}.
\bibitem[{Filipovi\v{c} et~al.(2017)Filipovi\v{c}, Petrovi\v{c}, and
  Benkner}]{filipovic2017autotuning}
\bibinfo{author}{J.~Filipovi\v{c}}, \bibinfo{author}{F.~Petrovi\v{c}},
  \bibinfo{author}{S.~Benkner},
\newblock \bibinfo{title}{Autotuning of {OpenCL} kernels with global
  optimizations},
\newblock in: \bibinfo{booktitle}{Proceedings of the 1st Workshop on AutotuniNg
  and aDaptivity AppRoaches for Energy Efficient HPC Systems ({ANDARE '17})},
  \bibinfo{year}{2017}.
\bibitem[{Nugteren(2018)}]{nugteren2018clblast}
\bibinfo{author}{C.~Nugteren},
\newblock \bibinfo{title}{{CLBlast}: A tuned {OpenCL} {BLAS} library},
\newblock in: \bibinfo{booktitle}{Proceedings of the International Workshop on
  {OpenCL}}, IWOCL '18, \bibinfo{publisher}{ACM}, \bibinfo{year}{2018}, pp.
  \bibinfo{pages}{5:1--5:10}. \DOIprefix\doi{10.1145/3204919.3204924}.

\end{thebibliography}

\end{document}